\documentclass[11pt]{article}

\usepackage{amsmath}
\usepackage{amssymb}
\usepackage{hyperref}

\usepackage{xcolor}
\usepackage{graphicx}
\usepackage{bbm}
\usepackage{maybemath}

\usepackage{cite}

\definecolor{garrosgreen}{rgb}{0.1, 0.4, 0.1}
\definecolor{dartmouthgreen}{rgb}{0.05, 0.5, 0.06}
\definecolor{duelferred}{rgb}{0.7, 0.2, 0.1}
\definecolor{cambridgeblue}{rgb}{0.1, 0.3, 1.0}
\definecolor{oxfordblue}{rgb}{0.05, 0.2, 0.7}

\newcommand{\ii}{{\mathrm i}}
\newcommand{\dd}{{\mathrm d}}

\newcommand{\calL}{\mathcal{L}}

\newcommand{\GeV}{\mathrm{GeV}}
\newcommand{\PeV}{\mathrm{PeV}}

\newcommand\del                 {\delta}
\newcommand\avp[1]              {|\vec{p}_{#1}|}

\newcommand{\checkC}{$\breve{\rm C}$}

\begin{document}

\title{Squeezing the Parameter Space for 
Lorentz Violation in the Neutrino Sector by Additional Decay Channels}

\author{Ulrich D. Jentschura\\
\small
Department of Physics, Missouri University\\
\small
of Science and Technology, Rolla, Missouri 65409, USA; ulj@mst.edu\\
\small
MTA--DE Particle Physics Research Group,\\
\small
P.O.~Box 51, H--4001 Debrecen, Hungary\\
\small
MTA Atomki, P.O.~Box 51, H--4001 Debrecen, Hungary}

\maketitle

\begin{abstract}
The hypothesis of Lorentz violation in the neutrino sector has intrigued
scientists for the last two to three decades. A number of theoretical
arguments support the emergence of such violations first and foremost for
neutrinos, which constitute the ``most elusive'' and ``least interacting''
particles known to mankind. It is of obvious interest to place stringent
bounds on the Lorentz-violating parameters in the neutrino sector.
In the
past, the most stringent bounds have been placed by calculating the
probability of neutrino decay into a lepton pair, a process made
kinematically feasible by Lorentz violation in the neutrino sector, above
a certain threshold.  However, even more stringent bounds can be placed on
the Lorentz-violating parameters if one takes into account, additionally,
the possibility of neutrino splitting, i.e., of neutrino decay into a
neutrino of lower energy, accompanied by ``neutrino-pair \checkC{}erenkov
radiation''. This process has negligible threshold and can be used to
improve the bounds on Lorentz-violating parameters in the neutrino sector.
Finally, we take the opportunity to discuss the relation of Lorentz and
gauge symmetry breaking, with a special emphasis on the theoretical models
employed in our calculations.
\end{abstract}


%
% Introduction
%
\section{Introduction}
\label{sec1}

Neutrinos are the most elusive particles within the 
Standard Model of Elementary Interactions.
Speculation about their tachyonic nature
started with Ref.~\cite{ChHaKo1985},
and has led to the development 
of a few interesting scenarios~\cite{JeNaEh2017}.
Within the Lorentz-violating 
scenarios~\cite{CoKo1998,Ko2004grav,KoMe2009,DiKoMe2009,KoMe2012,KoMe2013,%
DiKoMe2014,Di2014,Ta2014,StEtAl2015,Li2015},
many different tensor structures involving Lorentz-violating
parameters can be pursued. Here, we
assume that, in a preferred particular (observer) 
Lorentz frame, an isotropic dispersion relation
of the form  $E = \sqrt{ \vec p^2 \, v^2 + m^2}$ with $v > 1$ holds. 
(In this article, we use physical units
with $\hbar = \epsilon_0 = c = 1$.)
Formalizing the Lorentz-violating ideas,
the Lorentz--Violating Extension of the Standard Model (SME)
was developed with a strong inspiration from 
string theory~\cite{KoSa1989,KoPo1991}.

Kinematically, decay among neutrino mass eigenstates 
is allowed due to their mass differences,
while decay rates for ``ordinary'' neutrinos
within the Standard Model formalism
(for both Dirac as well as Majorana)
exceed the lifetime of Universe by orders of magnitude.
Lorentz-violating neutrinos undergo stronger
decay and energy loss mechanisms than ``ordinary'' neutrinos
because of their dispersion relation 
$E = \sqrt{ \vec p^2 \, v^2 + m^2} \approx |\vec p| \, v$ 
(at high energy), which makes a number of decay
channels (without GIM suppression,
see Refs.~\cite{PaWo1982,GlIlMa1970}) kinematically possible.

Some remarks on the origin of modified dispersion
relations of the form $E = \sqrt{ \vec p^2 \, v^2 + m^2} \approx 
|\vec p| \, v$ might be in order.
Modified dispersion relations could in principle be induced
by modified theories of gravity, and modifications
of the Einstein equivalence principle.
In Ref.~\cite{ACEtAl2011}, it has been suggested that the invariant arena for nonquantum
physics is a phase space rather than spacetime, and the locality of an even in
space-time is replaced by {\em relative locality} in which different observers
see different spacetimes, and the spacetimes they observe are energy and
momentum dependent. This hypothesis can lead to modified
dispersion relation of the kind investigated here.  
Effects due to quantum gravity may also induce modified
dispersion relations~\cite{PfEtAl2015,ToAnMi2019}. 
In a different context, Lorentz breaking induced at the Planck scale may
also induce such relations~\cite{AC2002,MaSm2004,GiLiSi2007}, 
in the sense of ``doubly special relativity'' 
which works on the assumption that dynamics are governed
by two observer-independent quantities, the speed of light and an additional
constant energy scale, which could be the Planck
energy scale (see also Ref.~\cite{MaSm2002}).
The phenomenological consequences of tiny Lorentz violations, which are
rotationally  and  translationally  invariant  in  a preferred  frame, and are
renormalizable while preserving anomaly cancellation and gauge invariance under
the Standard Model gauge group $SU(3) \otimes SU(2) \otimes U(1)$, have been
analyzed in Ref.~\cite{CoGl1999}.

There are a number of phenomena which could direct us to have a look 
at the neutrino sector for Lorentz violation.
Namely, for example, the early arrival of neutrinos from the 1987 supernova
still inspires (some) physicists.
Specifically, under the Mont Blanc, in the early morning hours of February 23, 1987,
a shower of neutrinos of interstellar origin arrived about 6 hours
earlier then the visible light from the Siderius Nuntius SN1987A
supernova. This event has been recorded in Ref.~\cite{DaEtAl1987},
and it was asserted that such an event could happen by accident
once in about 1000~years.
Direct measurements of neutrino velocities have
given results that are consistent with the speed
of light within experimental uncertainty,
but with the experimental result being a littler larger
than the speed of light.
For example, the MINOS experiment~\cite{AdEtAl2007minos} has measured
superluminal neutrino propagation velocities
which differ from the speed of light by a relative factor of
$(5.1 \pm 2.9) \times 10^{-5}$ at an energy of
about $E_\nu \approx 3 \, \GeV$,
compatible with an earlier FERMILAB experiment~\cite{KaEtAl1979}.
Furthermore, neutrinos cannot be used to transmit information (at least
not easily) because of their small interaction cross sections.
Superluminality of neutrinos would thus not necessarily lead to violation of
causality at a macroscopic level, 
as demonstrated in Appendix A.2 of Ref.~\cite{JeEtAl2014}.
Similar arguments have been made in Ref.~\cite{KoLe2001},
where it was shown that problems with microcausality,
in Lorentz-violating theories, are alleviated for 
small Lorentz-violating parameters and in so-called 
{\em concordant frames} where the boost velocities are 
not too large. For neutrinos, corresponding problems 
are further alleviated by the fact that their interaction
cross sections are small; hence, it becomes very hard to 
transport information superluminally even if the 
dispersion relation indicates such effects
(see also Appendix A.2 of Ref.~\cite{JeEtAl2014}).

We should also note that, when Lorentz invariance is violated, superluminality
does not necessarily lead to problems.  Under certain additional assumptions,
causality arguments related to superluminal signal propagation simply do not
apply in this case.  For example, if photons themselves propagate faster than
$c$ (the limiting velocity of massive standard fermions) in some
Lorentz-violating theories, then, as long as information propagates along or
inside the modified lightcone (as defined by a modified effective metric), it
can propagate faster than $c$ without implying causality issues (see, e.g., the
paragraph around Eq.~(9) in  Ref.~\cite{DuEtAl2008}).  Further information on
this point can be found in Refs.~\cite{AdKl2001,Re2010,KlSc2011,Sc2012,Sc2014}.
E.g., Ref.~\cite{AdKl2001} provides information on (micro)causality problems
for the purely timelike case of Maxwell-Chern-Simons theory (operator of
dimension 3 for photons).

Some more explanatory remarks on gauge invariance and Lorentz violation are 
probably in order.
One might argue that gauge invariance should be seen as the guiding principle
to make consistent a nonabelian gauge interaction mediated by spin-one massive
particles through the (experimentally confirmed) Higgs mechanism, and that
gauge invariance should be retained at all cost, even if Lorentz symmetry is
broken.  However, this demand overlooks two aspects.  The first is that the
origin of the Lorentz-violating terms could be assumed to be commensurate with
the Planck scale~\cite{KoSa1989,MaSm2002,AC2002},
in which case it is questionable if our usual 
concepts of gauge invariance could be transported without changes to the
extreme energy scales; in any case, we would assume the Standard Model gauge
group, at the energy scale relevant to Lorentz violation, to be replaced by a
unified gauge group, e.g., SO(1,13) 
(see Refs.~\cite{MB2001,MBNi2003,MB2015,MBNi2016,MB2017}).  Further fundamental modifications
of the gauge principle at extreme energy scales are also conceivable.  The
second and more important overlooked aspect is that Lorentz violation {\em
constitutes} a form of gauge invariance violation. Namely, Einstein's theory of
general relativity {\em is} the classical, gauged theory of gravitation, in
which the global Lorentz symmetry is elevated to a gauged, local symmetry~\cite{Bo2011}.
This observation offers the construction principle for the spin connection of
Dirac fields in curved space-times, which has been explored in recent
monographs~\cite{Bo2011} and
papers~\cite{JeNo2013pra,Je2018geonium}. To reemphasize the point, we recall that
Lorentz invariance violation {\em is} gauge invariance violation within General
Relativity~\cite{Bo2011}.

We organize the paper as follows. 
After a consideration of the calculation of the threshold for the 
superluminal decay processes (Sec.~\ref{sec2}), 
we outline the calculation of the decay and energy loss rates 
in Sec.~\ref{sec3}, before discussing an attractive 
scenario for Lorentz violation in the neutrino sector, and its 
signatures, in Sec.~\ref{sec4}. Conclusions are reserved for Sec.~\ref{sec5}.

%
% Figure 1
%
\begin{figure}[t!]%
\begin{center}
\begin{minipage}{0.8\linewidth}
\begin{center}
\begin{center}
\includegraphics[width=0.8\linewidth]{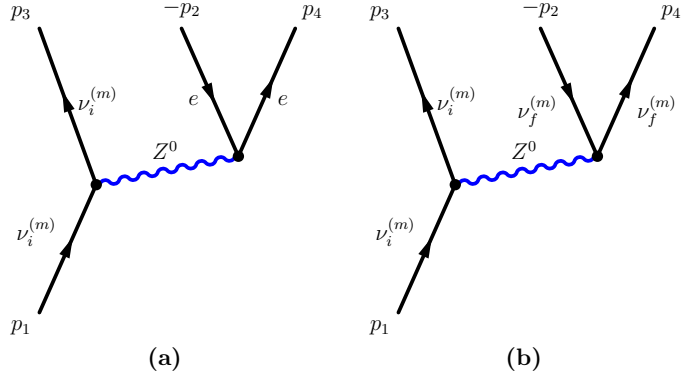}
\end{center}%
\caption{\label{figg1}%
In the lepton-pair \checkC{}erenkov radiation process~(a),
an oncoming Lorentz-violating initial neutrino mass eigenstate
$\nu_i^{(m)}$ decays, under emission of a 
virtual $Z^0$ boson, into an electron-positron pair.
The sum of the outgoing pair momenta is $p_2 + p_4$;
one observes the inverted direction of the 
fermionic antiparticle line.
The arrow of time is from bottom to top.
The (blue) bosonic line carries the four-momentum $q$.
Diagram~(b) describes the neutrino-pair \checkC{}erenkov radiation process,
with a final neutrino mass eigenstate $\nu_f^{(m)}$.}
\end{center}
\end{minipage}
\end{center}
\end{figure}%

%
% Threshold Considerations
%
\section{Threshold Considerations}
\label{sec2}

We refer to the lepton-pair 
\checkC{}erenkov radiation process (LPCR) in Fig.~\ref{figg1}(a) and the 
neutrino-pair \checkC{}erenkov radiation process 
(NPCR) depicted in Fig.~\ref{figg1}(b).
In order to make neutrino decay kinematically 
possible, it is necessary to fulfill certain threshold
conditions. Let us denote the outgoing 
fermions in the generic decay processes depicted 
in Fig.~\ref{figg1} by
\begin{equation}
\nu \to \nu + f + {\bar f} \,,
\end{equation}
with a pair of a massive fermion $f$ and its antiparticle
$\bar f$ being emitted in the process.

Energy-momentum conservation implies that 
(in the notation of Fig.~\ref{figg1})
\begin{equation}
(p_1 - p_3)^2 = q^2 = (p_2 + p_4)^2 \,.
\end{equation}
Let us first consider the case of 
a massive outgoing pair $2+4$, with rest mass $m_f$,
and vanishing Lorentz-violating parameter.
Threshold is reached for collinear emission 
geometry. The incoming four-momentum is 
$p_1 = (E_1, \vec p_1)$,
where $E_1 = \sqrt{\vec p_1^{\,2} \, v_i^2 + m_\nu^2} \approx 
| \vec p_1 | v_i$,
and $m_\nu$ is the neutrino mass.
Assuming an incoming neutrino energy well above its rest mass,
we can do this approximation.
By contrast, one has $p_3 = (0, \vec 0)$, so that the total 
four-momentum transfer $q$ goes into the pair.

A few remarks on the dispersion relations used in the current paper,
are in order. We observe that the relation $E_1 = \sqrt{\vec p_1^{\,2} \, v_i^2 + m_\nu^2}$ is 
isotropic, and it induces superluminal velocities.  However, it should be noted that the
modified dispersion relation applied to neutrino sector can introduce a rich
phenomenology even in other sectors, as oscillations, without requiring superluminal
velocities for $v_i > 1$. Many models can be explored
(see Refs.~\cite{ArEtAl2007,KoMe2009,DiKoMe2009,KoMe2012,%
DiKoMe2014,Li2013,AnMiTo2018,To2020}).
As explained above, we here assume a simple form for the modified dispersion relation, 
applicable at least in a preferred observer
Lorentz frame.

Returning to the discussion of the threshold, we observe that,
for collinear geometry, one has 
$p^\mu_2 = p^\mu_4 = (E_f, \vec p_f)$, where 
$E_f = \sqrt{ \vec p_f^{\,2} + m^2}$.
Under these assumptions,
$p_2 + p_4 = (2  \sqrt{ \vec p_f^{\,2} + m^2}, 2 \vec p_f)$, 
so that $(p_2 + p_4)^2 = 4 m_f^2$.
The threshold condition becomes
\begin{equation}
\label{thr1}
p_1^2 (v_i^2 - 1) \geq 4 m_f^2 \,,
\qquad
p_1 \approx E_1 \geq \frac{2 m_f}{\sqrt{v_i^2 - 1}} = 
\frac{2 m_f}{\sqrt{\delta_i}} \,.
\end{equation}
Here, 
\begin{equation}
\label{defdelta}
v_i = \sqrt{1 + \delta_i} \,.
\end{equation}
The threshold condition $E_{\rm th} = 2 m_f/\sqrt{\delta_i}$ 
has been used extensively in Refs.~\cite{CoGl2011,BeLe2012,SoNaJe2019}.
The formula~\eqref{thr1} implies that the 
threshold for NPCR is lower by at a least six orders 
of magnitude as compared to LPCR.

The kinematic considerations are very different in the high-energy
regime, when {\em both} the incoming (decaying) particle
as well as the outgoing particles are Lorentz violating.
Masses can be neglected.
In this case, one has at threshold
$p_2 = p_4 = (E_f, \vec p_f)$, where $E_f = | \vec p_f | \, v_f$,
so that at threshold
\begin{equation}
p_1^2 (v_i^2 - 1) \geq 4 p_f^2 (v_f^2 - 1) \,,
\end{equation}
Due to equipartition of the energy among each particle of the outgoing pair 
at threshold, one has
$p_f \approx E_f = E_1/2 \approx p_1/2$.
In this case, the threshold condition reduces to
\begin{equation}
(v_i^2 - 1) \geq(v_f^2 - 1) \,,
\qquad
\delta_i > \delta_f \,.
\end{equation}
Here, $v_f = \sqrt{1 + \delta_f}$.
For $\delta_i = \delta_f$, no phase space is available 
in order to accommodate for the decay.
This consideration explains why all results
communicated in Ref.~\cite{SoNaJe2019} display 
a factor $\delta_i - \delta_f$;
decay takes place from ``faster'' to ``slower''
mass eigenstates.

%
% Outline of the Calculation
%
\section{Outline of the Calculation}
\label{sec3}

The understanding of decay processes involving 
Lorentz-violation has been advanced through 
Refs.~\cite{CoGl2011,BeLe2012,SoNaJe2019}.
Let us briefly recall elements of the derivation 
given in Ref.~\cite{SoNaJe2019}.
One particular question is how to express the decay rate for an
(initially) flavor-eigenstate neutrino 
(the electroweak Lagrangian is flavor-diagonal)
in terms of mass eigenstates.
We have, in the same obvious notation as used
in Ref.~\cite{SoNaJe2019},
\begin{equation}
\nu^{(f)}_k = \sum_\ell U_{k \ell} \, \nu^{(m)}_\ell \,,
\end{equation}
with the Pontecorvo--Maki--Nakagawa--Sakata (PMNS) matrix $U_{k \ell}$.
The interaction interaction $\calL_W$ with the $Z^0$ boson in the 
flavor basis is
\begin{equation}
\calL_W = -\frac{g_w}{4 \, \cos\theta_W} \,
\sum_{k} \overline\nu^{(f)}_k \, \gamma^\mu (1 - \gamma^5) \, \nu^{(f)}_k  \, Z_\mu \,.
\end{equation}
Here, $g_w$ is the weak coupling constant,
and $\theta_W$ is the Weinberg angle.
A unitary transformation leads to
\begin{equation}
\calL = -\frac{g_w}{4 \, \cos\theta_W} \,
\sum_{k,\ell,\ell'} U^+_{\ell k} \, U_{k \ell'} \,
\overline\nu^{(m)}_\ell \, \gamma^\mu (1 - \gamma^5) \,
\nu^{(m)}_{\ell'}  \, Z_\mu \,.
\end{equation}
The interaction with the $Z^0$ boson in the 
mass eigenstate basis therefore reads as follows,
\begin{equation}
\label{lag_escalafon2}
\calL = -\frac{g_w}{4 \, \cos\theta_W} \,
\sum_{\ell}
\overline\nu^{(m)}_\ell \, \gamma^\mu (1 - \gamma^5) \,
\nu^{(m)}_{\ell}  \, Z_\mu \,.
\end{equation}
In order to model the free Lorentz-violating neutrino Lagrangian,
one introduces an effective metric with a tilde:
\begin{equation}
\calL = \sum_\ell \ii \, \overline \nu^{(m)}_\ell \,
\gamma^\mu \, (1-\gamma^5) \,
{\tilde g}_{\mu\nu}(v_\ell) \,
\partial^\nu \nu^{(m)}_\ell \,.
\end{equation}
The modified metric defines a modified light cone according to
${\tilde g}_{\mu\nu}(v_\ell) \, k^\mu \, k^\nu = 0$.
Here,
\begin{equation}
{\tilde g}_{\mu\nu}(v_\ell) = {\rm diag}(1, -v_\ell, -v_\ell, -v_\ell)  \,.
\end{equation}
The dispersion relation 
\begin{equation}
E_\ell = |\vec p| \, v_\ell 
\end{equation}
follows as the massless limit of 
$E_\ell = \sqrt{ (|\vec p| \, v_\ell)^2 + m_\ell^2}$.
For neutrinos, we know that the $m_\ell$ terms are different.
So, there is reason to assume that the $\delta_\ell = \sqrt{v_\ell^2 - 1}$
terms are also different among mass (flavor) eigenstates,
if they are nonvanishing.

One defines parameters $v_{\rm int}$ and $\delta_{\rm int}$ by the 
relation
\begin{equation}
v_{\rm int} = \sqrt{1 + \delta_{\rm int}}
\end{equation}
for the unified description of 
LPCR and NPCR; the effective four-fermion Lagrangian for the 
process reads as
\begin{align}
\calL_{\rm int} =& \;
f_e \, \frac{G_F}{2 \sqrt{2}} \,
\overline\nu^{(m)}_i \, \gamma^\lambda \, (1 - \gamma^5) \, \nu^{(m)}_i
{\tilde g}_{\lambda\sigma}(v_{\rm int}) \;
\bar{\psi}_f \, \gamma^\sigma \, (c_V - c_A \, \gamma^5) \, \psi_f \,.
\end{align}
Cohen and Glashow~\cite{CoGl2011} set $v_{\rm int} = 1$.
(In Ref.~\cite{SoNaJe2019}, on a number of 
occasions, the parameter used in Ref.~\cite{CoGl2011} 
had been inadvertently indicated as 
$v_{\rm int} = 0$, which is not the case. 
We take the opportunity to point out that of course,
the parameter $v_{\rm int} = 1$ implies that 
$\delta_{\rm int} = 0$, which was the intended statement 
in Ref.~\cite{CoGl2011}.)
Bezrukov and Lee~\cite{BeLe2012} use the parameters
$v_{\rm int} = 1$ (``model I'') and $v_{\rm int} = v_i$
(``model II''). In Ref.~\cite{SoNaJe2019}, the parameter
$v_{\rm int}$ is kept as a variable.
As explained in detail in Ref.~\cite{JeNaSo2019},
``gauge invariance'' (with respect to a restricted subgroup 
of the electroweak sector) can be restored if one uses
the value $v_{\rm int} = v_i \, v_f$.
Both Cohen and Glashow~\cite{CoGl2011}, as well as Bezrukov and Lee~\cite{BeLe2012},
assume that $\delta_f = 0$ for LPCR.
The parameter $f_e$ characterizes the process:
\begin{equation}
f_e = \left\{ \begin{array}{ll}
1, & \qquad \psi_f = \nu^{(m)}_f \\
2, & \qquad \psi_f = e \\ 
\end{array} \right. \,.
\label{eq:fe}
\end{equation}
Approximately, one has
\begin{equation}
(c_V, c_A) = \left\{ \begin{array}{ll}
(1,1) & \qquad \psi_f = \nu^{(m)}_f \\
(0,-\tfrac12), & \qquad \psi_f = e \\
\end{array} \right. \,.
\label{eq:cAcV}
\end{equation}
The characteristic matrix element is
\begin{align}
{\mathcal M} = f_e \frac{G_F}{2\sqrt{2}}
\left[\bar{u}_i(p_3)\gamma^\lambda(1-\gamma^5)u_i(p_1)\right]
{\tilde g}_{\lambda\sigma}(v_{\rm int})
\left[\bar{u}_f(p_4)(c_V \gamma^\sigma - c_A \gamma^\sigma \gamma^5)v_f(p_2)\right]\,.
\end{align}
Key to the calculation is the fact that one can split the 
three-particle outgoing phase space 
\begin{align}
\label{Gamma}
\Gamma =& \; \frac{1}{2E_1} \int \dd \phi_3(p_2,p_3,p_4;p_1)
\frac{1}{n_s} \sum_{\mathrm{spins}} |{\mathcal M}|^2
\nonumber\\[0.1133ex]
=& \; \frac{1}{2E_1} \int_{M^2_{\mathrm{min}}}^{M^2_{\mathrm{max}}} \frac{\dd M^2}{2\pi}
\dd \phi_2(p_3,p_{24};p_1) \,
\dd \phi_2(p_2,p_4;p_{24})
\frac{1}{n_s} \sum_{\mathrm{spins}} |{\mathcal M}|^2 \,.
\end{align}
with appropriate limits for $M^2_{\mathrm{min}}$
and $M^2_{\mathrm{max}}$ being given as follows,
\begin{equation} 
M^2_{\mathrm{min}} = \del_f (\avp{2} + \avp{4})^2\,,
\qquad
M^2_{\mathrm{max}} = \del_i (\avp{1} - \avp{3})^2\,.
\end{equation}
The following splitting relation for the 
phase space is crucial to a simplification of the 
integrations [for details, see Ref.~\cite{ByKa1973} and 
Eq.~(43) of Ref.~\cite{SoNaJe2019}],
\begin{equation}
\label{phase}
\begin{split}
& \dd \phi_3(p_2,p_3,p_4;p_1) \\
& =
\int \frac{\dd M^2}{2\pi}
\underbrace{\frac{\dd^4 {p_3}}{(2\pi)^3} \del_{+}(p_3^2 - \del_i k_3^2)
\frac{\dd^4 p_{24}}{(2\pi)^3} \del_{+}(p_{24}^2 - M^2)
(2\pi)^4 \del^{(4)}(p_1-p_3-p_{24})}_{= \dd \phi_2(p_3,p_{24};p_1)}
\\
& \times
\underbrace{
\frac{\dd^4 p_2}{(2\pi)^3} \del_{+}(p_2^2 - \del_f k_2^2)
\frac{\dd^4 p_4}{(2\pi)^3} \del_{+}(p_4^2 - \del_f k_4^2)
(2\pi)^4 \del^{(4)}(p_{24}-p_2-p_4)}_{=  \dd \phi_2(p_2,p_4;p_{24})}
\\
&
= \int \frac{\dd M^2}{2\pi} \dd \phi_2(p_3,p_{24};p_1) \dd \phi_2(p_2,p_4;p_{24})\,.
\end{split}
\end{equation}
Here,  $p_{24} = p_2 + p_4$, and we denote the 
spatial part of the four-vector $p_i^\mu$ as $\vec k_i$ with 
$i=1,2,3,4$, so that 
$p_i^\mu = (E_i, \vec k_i)$, and since $E_i = (1 + \delta_i) \, | \vec k_i|$,
one has $g^{\mu\nu} \, {\tilde g}_{\mu\alpha} \, {\tilde g}_{\nu\beta} \, p_i^\alpha p_i^\beta = 
g_{\mu\nu} p_i^\mu p_i^\nu - \delta_i \, k_i^2 = p_i^2 -  \delta_i \, k_i^2$
[see also Eq.~(40) of Ref.~\cite{SoNaJe2019}]. This relation explains the
argument of some of the Dirac-$\delta$ functions in Eq.~\eqref{phase}.
The general result for the decay rate, unifying both 
processes depicted in Fig.~\ref{figg1}, reads as follows,
\begin{equation}
\label{eq:Gamma-general}
\begin{split}
&
\Gamma_{\nu_i \to \nu_i \psi_f \bar{\psi}_f} = \frac{G_F^2 k_1^5}{192\pi^3}
        f_e^2 \frac{c_V^2+c_A^2}{420 n_s} (\delta_i-\delta_f) 
        \bigg[(60 - 43 \sigma_i)(\delta_i-\delta_f)^2
\\ & \quad\quad+
        2(50 - 32\sigma_i - 25\sigma_f + 7 \sigma_i \sigma_f)
        (\delta_i-\delta_f) \delta_f
\\ & \quad\quad+
        7(4 - 3\sigma_i - 3\sigma_f + 2 \sigma_i \sigma_f)
        \delta_f^2
        + 
        7 \delta_{\rm int}^2\bigg]\,.
\end{split}
\end{equation}
This result vanishes for $\delta_i = \delta_f$
(see the discussion in Sec.~\ref{sec2}).
Cohen and Glashow~\cite{CoGl2011}
 have $n_s = 2$ active spin states for the 
(initial) neutrino, while Bezrukov and Lee~\cite{BeLe2012} calculate with $n_s = 1$,
implying that the authors of   Ref.~\cite{BeLe2012} assume that only one spin state
exists in nature. This assumption affects the averaging over the initial quantum states
involved in the process.
The $\sigma$ parameters depend on the way in which 
spin polarization sums are carried out,
\begin{equation}
\sigma_i = \left\{ \begin{array}{ll}
0, & \quad \mbox{CG spin sum for $\nu_i$}\\
1, & \quad \mbox{BL spin sum for $\nu_i$}\\ 
\end{array} \right. , \;\;
\sigma_f = \left\{ \begin{array}{ll}
0, & \quad \mbox{CG spin sum for $\psi_f$}\\
1, & \quad \mbox{BL spin sum for $\psi_f$}\\ 
\end{array} \right. .
\end{equation}
In Ref.~\cite{CoGl2011},
the Cohen--Glashow (CG) spin sum (``polarization sum'') is simply taken as 
the standard spin sum for massless fermions,
\begin{equation}
\sum_s \nu_{\ell,s} \otimes \bar{\nu}_{\ell,s} = p^\mu g_{\mu\nu} \gamma^\nu \,.
\end{equation}
In Ref.~\cite{BeLe2012}, the
Bezrukov--Lee (BL) spin sum is based on a somewhat
more advanced treatment of the eigenspinors of
superluminal neutrino mass eigenstates and reads as
\begin{equation}
\label{BLspinsum}
\sum_s \nu_{\ell,s} \otimes \bar{\nu}_{\ell,s} = 
p^\mu {\tilde g}_{\mu\nu}(v_\ell) \gamma^\nu \,.
\end{equation}

The general result for the energy loss rate,
applicable to both processes in Fig.~\ref{figg1},
reads as
\begin{equation}
\begin{split}
&\frac{\dd E_{\nu_i \to \nu_i \psi_f \bar{\psi}_f}}{\dd x} 
= -\frac{G_F^2 k_1^6}{192\pi^3} 
f_e^2 \frac{c_V^2+c_A^2}{672 n_s} (\delta_i-\delta_f) 
\\ &\quad\times
\bigg[(75 - 53 \sigma_i)(\delta_i-\delta_f)^2
+ (122 - 77\sigma_i - 61\sigma_f + 16 \sigma_i \sigma_f)
(\delta_i-\delta_f) \delta_f
\\ & \quad\quad+
8(4 - 3\sigma_i - 3\sigma_f + 2 \sigma_i \sigma_f) \delta_f^2
+ 8 \delta_{\rm int}^2\bigg]\,.
\end{split}
\end{equation}
In Ref.~\cite{SoNaJe2019},
we have verified and checked compatibility 
with all formulas contained in Refs.~\cite{CoGl2011} and~\cite{BeLe2012}.
This is important because it confirms that the 
model dependence of the results is only contained
in the numerical prefactors, but not 
in the overall scaling of the results.

As outlined in Ref.~\cite{SoNaJe2019},
one can parameterize the results for NPCR as follows,
\begin{equation}
\label{NPCRfunc}
\Gamma_{\nu_i \to \nu_i \nu_f \bar{\nu}_f}
= b \frac{G_F^2}{192\pi^3} k_1^5\,,
\qquad
\frac{\dd E_{\nu_i \to \nu_i \nu_f \bar{\nu}_f}}{\dd x} =
-b' \frac{G_F^2}{192\pi^3} k_1^6\,.
\end{equation}
For the CG spin sum, one obtains the following
$b$ parameters,
\begin{subequations}
\label{resbCG}
\begin{align}
b_{\mathrm{CG}} =& \; \frac{1}{7} (\del_i - \del_f)
\left[(\del_i - \del_f)^2 + \frac{5}{3} \del_f (\del_i - \del_f)
+ \frac{7}{15}\del_f^2\right]\,,
\\
b'_{\mathrm{CG}} =& \; \frac{25}{224} (\del_i - \del_f)
\left[(\del_i - \del_f)^2 
+ \frac{112}{75} \del_f (\del_i - \del_f) + \frac{32}{75}\del_f^2\right]\,.
\end{align}
\end{subequations}
For the BL spin sum, one obtains
\begin{subequations}
\label{resbBL}
\begin{align}
b_{\mathrm{BL}} =& \; \frac{17}{210} (\del_i - \del_f)
\left[(\del_i - \del_f)^2 + \frac{7}{17}\del_{\rm int}^2\right]\,,
\\
b'_{\mathrm{BL}} =& \; \frac{11}{168} (\del_i - \del_f)
\left[(\del_i - \del_f)^2 + \frac{4}{11}\del_{\rm int}^2\right]\,.
\end{align}
\end{subequations}
Typically, one finds~\cite{SoNaJe2019} 
numerical prefactors in these formulas
are larger than those for 
LPCR by a factor of four or five. Also, NPCR has negligible threshold.

In papers of Stecker and Scully~\cite{St2014,StSc2014,StEtAl2015},
the following bound is
derived for the Lorentz-violating 
parameter of the electron-positron field alone
(watch out for a difference in the conventions used for defining
the $\delta_e$ parameter):
\begin{equation}
\label{deltae}
\delta_e \leq 1.04 \times 10^{-20} \,.
\end{equation}
We should stress that this bound concerns oncoming electrons (not neutrinos!)
and has nothing to do with the processes studied here.

The observation of very-high-energy neutrinos by 
IceCube, taking into consideration 
the LPCR process (but not NPCR!), implies that the Lorentz-violating 
parameter for neutrinos cannot be larger than 
(Ref.~\cite{StEtAl2015})
\begin{equation}
\label{deltanu}
\delta_\nu \leq 2.0 \times 10^{-20} \,.
\end{equation}
This bound is based on the assumption that $\delta_e$ and
$\delta_\nu$ are different.
Colloquially speaking, we can say that, 
if $\delta_\nu$ were larger, then ``Big Bird'' 
(the 2\,PeV specimen found in IceCube,
see Refs.~\cite{AaEtAl2013icecube,AaEtAl2014icecube})
would have already decayed before it arrived at the
IceCube detector.
However, the full analysis requires Monte Carlo 
simulations involving astrophysical 
data and is much more involved~\cite{St2014,StSc2014,StEtAl2015}.

Provided the Lorentz-violating parameters for the 
different neutrino mass eigenstates
are different, low-energy neutrinos
are affected by the decay and energy loss
processes connected with NPCR,
in view of a negligible threshold for NPCR.
As already emphasized, typical 
numerical coefficients for NPCR are a factor 
of four or five larger than for LPCR, depending on the 
model used for the spin sums.
This enhances the importance of the NPCR effect.
Inspired by Eq.~\eqref{deltanu},
we thus conjecture here that a full analysis of astrophysical
data, using the NPCR process as a limiting factor 
for the observation of high-energy neutrinos,
should yield a bound on the order of
\begin{equation}
\label{deltaconj}
| \delta_i - \delta_f| \leq \frac{1}{5^{1/3}} \times 2.0 \times 10^{-20} 
\sim 1.2 \times 10^{-21} \,,
\end{equation}
where the prefactor takes into account the 
scaling of the effect with the $\delta$ parameter.
Specifically, the decay and energy loss rates
typically scale with the factor $(\delta_i - \delta_f)^3$.
It would be very fruitful if this conjecture were to be checked against 
astrophysical data in an independent investigation.

%
% An Attractive Scenario
%
\section{An Attractive Scenario}
\label{sec4}

At first, one might see a dilemma:
Within a fully $SU(2)_L$ gauge-invariant theory,
with uniform Lorentz-violating parameters over all particle generations,
one necessarily has $\delta_\nu = \delta_e$
(see Ref.~\cite{JeNaSo2019} for a detailed discussion),
and so, the bound $\delta_\nu \leq 2.0 \times 10^{-20}$
given in Eq.~\eqref{deltanu}
is not applicable, because the LPCR process does not
exist. But then, one has to acknowledge that the 
bound $\delta_e \leq 1.04 \times 10^{-20}$
given in Eq.~\eqref{deltae}, which 
is originally derived for electrons, based on other physical processes,
automatically also applies to the neutrino sector.

So, the dilemma is that either, one has to give 
up gauge invariance and use different Lorentz-violating parameters for 
neutrinos as opposed to charged leptons within the same particle generation,
or, if one insists on gauge invariance, 
or, assume different Lorentz-violating parameters for different
generations. 
Otherwise, the insistence on gauge invariance would  defeat
part of the purpose of looking at the neutrino
sector for Lorentz violation.
This is because in the latter case, 
for uniform Lorentz-violating parameters among all three generations,
because, by assumption, the Lorentz-violating parameters for neutrinos
and charged left-handed leptons within the 
same $SU(2)_L$ doublet are necessarily the same,
the tight bounds on Lorentz-violating 
parameters in the charged-fermion sector
automatically apply to the neutrino sector as well\footnote{We here ignore the somewhat remote possibility of 
different Lorentz violating parameters for the right-handed
and left-handed sectors of one and the same generation.}. 
This observation has important consequences when 
examining the first-generation 
$SU(2)_L$ doublet, consisting of $(\nu_e, e_L)$.
Electrons and positrons are 
stable particles, and small violations of Lorentz 
invariance would lead to 
violations of causality on a macroscopic level
(see Appendix A.2 of Ref.~\cite{JeEtAl2014})
\footnote{Note, also,
that this statement does not hold if the limiting velocity for fermions 
turns out to be smaller as opposed to larger than the speed of light.
Furthermore, there are conceivable modifications of 
Maxwell theory, as already discussed in Sec.~\ref{sec1},
where causality violations are avoided due to 
modified light cones~\cite{AdKl2001,Re2010,DuEtAl2008,KlSc2011,Sc2012,Sc2014}.}.
Conversely, if we had to carry over all restrictions on Lorentz-violating 
electron parameters to the electron neutrino 
sector, then this would nullify all the motivations
listed in Sec.~\ref{sec1} for investigating the 
first-generation neutrino sector.

On the contrary,
If one accepts the necessity that 
different Lorentz-violating parameters should be used for 
each of the three known particle generations,
then one needs to acknowledge that the parameter space for 
differential Lorentz-violation among 
neutrino mass eigenstates is restricted by additional
constraints due to the NPCR process~\cite{SoNaJe2019}.
An attractive gauge-invariant scenario could still be 
found, as follows. 
Namely, one might observe that, as per the discussion
in Appendix A.2 of Ref.~\cite{JeEtAl2014},
causality violations due to Lorentz violation
are less severe for unstable particles,
which decay and therefore are not amenable to 
the reliable transport of information.
Part of the above sketched dilemma could thus 
be avoided as follows.
One first observes that, as per the above argument,
problems with respect to causality are 
less severe in the second-generation $SU(2)_L$ 
doublet $(\nu_\mu, \mu_L)$ and also in the 
third-generation $SU(2)_L$ doublet $(\nu_\tau, \tau_L)$,
which are composed entirely of unstable particles.
Full gauge invariance can be retained if we 
assume generation-dependent Lorentz-violating parameters
$\delta_e$, $\delta_\mu$, and $\delta_\tau$, 
for the three $SU(2)_L$ doublets, 
which could be encoded in modified Dirac matrices
$\widetilde \gamma^i = v_f \, \gamma^i $
with $f=e,\mu,\tau$ [see Eq. (5) of Ref.~\cite{JeNaSo2019}].
In the charged-fermion sector, we have nearly 
no mixing of mass and charge eigenstates.
Let us then go into the high-energy regime where,
where mass and flavor eigenstates, under the 
assumptions
\begin{equation}
\delta_\mu, \delta_\tau > 0 \,,
\qquad
\delta_\mu \neq \delta_\tau \,,
\qquad
\delta_e = 0 \,,
\end{equation}
become equal.
In this case, at high energy, one would have two 
neutrino mass eigenstates, which asymptotically 
approach the muon neutrino and tau neutrino flavor eigenstates
at very high energy,
decay into electron-positron pairs and (asymptotically)
electron neutrinos, via LPCR and NPCR.

Of course, other scenarios and flavor and mass mixing
phenomenologies are also possible,
as discussed in Sec.~IV B of Ref.~\cite{SoNaJe2019}.
In general, one could interpret the emergence of a 
specific predominant flavor composition of incoming
super-high-energy cosmic neutrinos, consistent 
with one, and only one, specific mass eigenstate,
as a signature of Lorentz violation. This is because a 
single, defined, oncoming
mass eigenstate would be consistent with the two other
mass eigenstates being ``faster'' and thus decaying into the 
single ``slow'' eigenstate.

In all discussed scenarios, one might find a conceivable
explanation for the apparent cutoff in the cosmic 
neutrino spectrum at about $2\, \PeV$,
at the expense of reducing the allowed regime 
of Lorentz-violating $\delta$ parameters to 
the range of about $10^{-20}$.
In our ``attractive scenario'',
one retains gauge invariance as outlined in 
Sec.~4 of Ref.~\cite{JeNaSo2019} and 
still is able to account for a super-high-energy cutoff 
of the cosmic neutrino spectrum.
Experimental confirmation or dismissal of this hypothesis
will require better cosmic 
neutrino statistics at very high energies.

%
% Conclusions
%
\section{Conclusions}
\label{sec5}

The existence of the NPCR process [see Fig.~\ref{figg1}(b)]
reveals a certain dilemma for Lorentz-violating neutrinos (provided 
the Lorentz violating parameters indicate superluminality).
Namely, under the hypothesis of a nonvanishing
Lorentz-violating parameter $\delta$, given as in 
Eq.~\eqref{defdelta}, the virtuality 
\begin{equation}
E^2- p^2 = p^2 (v^2 -1 )
\approx E^2 (v^2 -1 )= E^2 \, \delta 
\end{equation}
of a neutrino becomes large for large energy, rendering 
a number of decay processes kinematically
possible. Conversely, based on high-energy 
astrophysical observations, very strict bounds
can be imposed on the Lorentz-violating 
parameters [see Eqs.~\eqref{deltae},~\eqref{deltanu},
and~\eqref{deltaconj}].

Deep connections exist between Lorentz violation and 
gauge invariance. In Ref.~\cite{ChJe2008},
it is shown that spontaneous Lorentz violation 
can lead to an effective low-energy field theory 
with both Lorentz-breaking as well as
gauge-invariance breaking terms.
According to Refs.~\cite{He1957,Bj1963,BB1963,Eg1976,%
ChFrNi2001prl,ChFrNi2001npb,Bj2001,AzCh2006,ChJe2007,ChFrJeNi2008,ChJe2008,ChFrNi2009},
even the photon could potentially be formulated as
the Nambu-Goldstone boson linked to spontaneous Lorentz invariance violation.
(This {\em ansatz} was originally formulated before
electroweak unification.)
For a broader view of this point, we refer
to Appendix~A of Ref.~\cite{JeNaSo2019}.
If one insists on the persistence of gauge
invariance within the electroweak sector,
then one has to acknowledge that bounds on 
Lorentz-violating parameters for charged leptons
[e.g., Eq.~\eqref{deltae}] also apply to the 
neutrino sector [thus lowering the bound otherwise
given in Eq.~\eqref{deltanu} by a factor two,
and further restricting the available parameter
space for Lorentz-violating parameters in the
neutrino sector].
Also, the assumption that $\delta_\nu = \delta_e$ 
would defeat the purpose of looking at neutrinos 
for Lorentz violation.
If one insists on gauge invariance and still 
pursues the exploration of Lorentz violation in the 
neutrino sector, then more sophisticated considerations
are required (see Sec.~\ref{sec4}).  Namely, one could potentially
invoke flavor-dependent differential Lorentz violation across 
generations (i.e., with different Lorentz-violating parameters 
for each generation). In this case, flavor and mass eigenstates
would become identical in the high-energy limit,
and decay and energy loss processes could potentially 
contribute to an explanation for the apparent cutoff
in the cosmic neutrino spectrum in the range of a 
few $\PeV$ (see Refs.~\cite{AaEtAl2013icecube,AaEtAl2014icecube}
and the discussion in Sec.~\ref{sec4}).

\section*{Acknowledgments}

The authors acknowledges support from the
National Science Foundation (Grant PHY--1710856)
as well as insightful conversations with G.~Somogyi and I.~N\'{a}ndori.

\end{document}